\documentclass[prl,twocolumn,floatfix]{revtex4}
\usepackage{graphicx}
\usepackage{amssymb}

\begin{document}
\title{A junction of three quantum wires: restoring time-reversal 
symmetry by interaction}
\author{X.\ Barnab\'e-Th\'eriault}\thanks{Xavier B.-T. passed away in a tragic 
traffic accident on August 15, 2004.}
\author{A.\ Sedeki}
\author{V.\ Meden}
\author{K.\ Sch\"onhammer}
\affiliation{Inst.\ f.\ Theoret.\  Physik, Universit\"at G\"ottingen, 
Friedrich-Hund-Platz 1, D-37077 G\"ottingen, Germany}

\begin{abstract}
We investigate transport of correlated fermions through a 
junction of three one-dimensional quantum wires pierced by a 
magnetic flux. 
We determine the flow of the conductance 
as a function of a low-energy cutoff in the entire 
parameter space. For attractive interactions 
and generic flux the fixed point with maximal asymmetry of the 
conductance  is the stable one, as conjectured recently. For 
repulsive interactions and arbitrary flux we find a
line of stable fixed points with vanishing conductance 
as well as stable fixed points with  symmetric conductance 
$(4/9)(e^2/h)$. 
\end{abstract}
\maketitle     

Electronic transport through quasi one-dimensional (1d) 
systems is of current experimental and theoretical
interest. In one spatial dimension correlations play 
an important role and the physical properties of interacting 
fermions show distinctive non-Fermi liquid
features. Generically such systems can be described as
Tomonaga-Luttinger liquids (TLLs) characterized by a
vanishing quasi-particle weight and power-law scaling of 
correlation functions.\cite{KS}  
For spinless fermions, on which we focus here, the
characteristic exponents are expressible in terms of the 
interaction dependent TLL parameter $K$. 
For repulsive interactions $0< K <1$, while $K>1$ in the attractive
case. In TLLs inhomogeneities have a dramatic effect 
as can be inferred from the singular behavior of 
response functions of homogeneous models.\cite{LutherPeschel,Mattis} 

In an important first step the conductance $G$ through a TLL with a 
{\it single impurity} was understood.\cite{KaneFisher,Fendley}. For
vanishing energy scale $s$ (e.g.\ temperature) 
and $0<K<1$, $G$ tends towards 0 following a 
power-law. The low-energy 
physics is governed by the  ``decoupled chain'' fixed point (FP). The scaling 
dimension of a hopping term connecting two open ends is $1/K$ and leads 
to $G \propto s^{2 (1/K-1)}$. 
One can understand this behavior in a simple picture. 
Due to the interaction the self-energy develops long range oscillatory
behavior and the scattering off the resulting effective impurity 
potential leads to the power-law suppression of
$G$.\cite{MatveevGlazman,VM1,Tilman}  
For $K>1$, the conductance approaches the impurity free limit. 
Close to the ``perfect chain'' FP an impurity has 
scaling dimension $K$ and the correction to the impurity free 
conductance scales as $s^{2(K-1)}$. In this case the
effective impurity potential leads to a resonance at the chemical
potential.  

Recently junctions of several quasi 1d quantum wires were realized 
experimentally with single-walled carbon 
nanotubes.\cite{Fuhrer,Terrones} They might form the basis of electronic 
devices. 
Already the physics of the three wire junction (Y-junction) is 
considerably richer than the one of a single impurity.
Taking into account correlations transport
through such systems was investigated  
theoretically in Refs.\ \onlinecite{Nayak} and \onlinecite{Claudio}. 
These studies posed a number of interesting questions. 
In Ref.\ \onlinecite{Claudio} a symmetric triangular 
Y-junction pierced by a magnetic flux $\phi$ 
(measured in units of the flux quantum $h c/e$) was studied. In this
geometry time reversal symmetry is broken. In the 
non-interacting case this generically leads to an asymmetry of the 
conductance from wire $\nu$ to wire $\nu'$ and vice versa: $G_{\nu,\nu'} \neq
G_{\nu',\nu}$. For $K>1$ at  
flux $\phi=\pm \pi/2$  one FP was found 
applying an exact method adopted from boundary 
conformal field theory. The FP corresponds to the case of {\it maximal
asymmetry} of $G_{\nu,\nu'}$. We here consider TLL wires 
connected to semi-infinite Fermi liquid (FL) leads with TLL-FL
contacts that are modeled to be free of fermion 
backscattering. 
The results of Ref.\ \onlinecite{Claudio} 
obtained for semi-infinite TLLs can be extended to our 
modeling. Then maximal asymmetry is given by 
$G_{\nu,\nu'}=1 (0)$ and $G_{\nu',\nu}=0 (1)$ in units of the conductance 
quantum $e^2/h$.\cite{Claudioprivate} 
The scaling dimension of the most relevant operator at this 
FP is $\Delta=4K/(3+K^2)$ and the correction to the FP  
conductance scales as $s^{2(\Delta -1)}$.\cite{Claudio} 
For $1 < K < 3$, $2(\Delta  -1) > 0$ and the ``maximal asymmetry'' 
FP is attractive. This implies that independently of the 
junction parameters at low energy scales the conductance is 1 
from $\nu$ to $\nu'$ and 0 from $\nu'$ to $\nu$ or vice versa.
It was {\it conjectured} that for $1 < K< 3$ 
and {\it all} $\phi$, except $|\phi|/\pi$ being integer, this FP 
is the only stable one. 

We investigate the same physical problem considering 
{\it arbitrary fluxes} and attractive as well as {\it repulsive 
interactions.}
An approximation scheme  based on the functional 
renormalization 
group (fRG) method is used. It was earlier applied successfully to 
transport problems in inhomogeneous TLLs. 
By comparison with numerical results and 
exact scaling relations this scheme was shown to be reliable 
for $1/2 \leq  K \leq 3/2$. 
In particular, the scaling dimensions 
of the two FPs of the single impurity problem come out correctly  
to leading order in the interaction $U$.\cite{VM1,Tilman} 
For the  Y-junction we here 
confirm the conjectured stability of the ``maximal asymmetry'' FP  
for {\it all} $\phi$, with $|\phi|/\pi$ not integer, 
and reproduce the scaling dimension $\Delta$ to 
leading order in $U$. For $U>0$ and arbitrary $\phi$ a line of stable 
FPs with $G=0$ is identified.
In cases with symmetric conductance we suppress the indices on $G$.
We surprisingly find additional ``perfect junction'' FPs with 
{\it symmetric} conductance $G=4/9$ that for $U>0$ are {\it stable}
and have a scaling dimension not discussed before.
In a Y-junction of non-interacting wires without flux  
$G=4/9$ is the value maximally allowed by symmetry.\cite{Nayak}
Although time reversal symmetry is explicitly broken,  
for systems that flow towards the ``perfect junction'' FPs 
the {\it electron correlations} cause the conductance -- an observable 
that is commonly believed to indicate this breaking of symmetry -- 
to behave as if the symmetry is restored. In a certain sense (see
below) this also holds for systems that flow towards the line of 
$G=0$ FPs and thus for all parameters. 

\begin{figure}[tbh]
\begin{center}
\includegraphics[width=0.16\textwidth,clip]{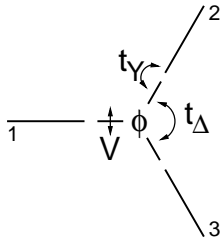}
\end{center}
\vspace{-0.6cm}
\caption[]{Symmetric junction of three quantum wires. \label{skizze}}
\end{figure}

Our wire Hamiltonian is   
\begin{eqnarray}
\label{ham0}
H_\nu & = & - \sum_{j=1}^{\infty} \left( c_{j,\nu}^\dag c_{j+1,\nu} +
  \mbox{h.c.} \right) \nonumber \\*
&& + \sum_{j=1}^{N-1} U_j \left(n_{j,\nu}- \frac{1}{2} \right)  
\left(n_{j+1,\nu}- \frac{1}{2}\right)      
\end{eqnarray}
in standard second-quantized notation. The different wires are
indicated by an index $\nu=1,2,3$. The amplitude of the nearest neighbor 
hopping is set to 1.
The nearest neighbor interaction $U_{j}$ is allowed to depend on 
the position. It is set to zero for $j > N$, i.e.\ the wires of length $N$ 
are connected to non-interacting leads. Close to the contacts the
interaction is switched off (spatially) smoothly to avoid any fermion
backscattering.\cite{Tilman} The bulk value of the interaction is
denoted by $U$. We here consider the half-filled band case. To prevent 
depletion of the interacting region we shifted the operator 
$n_{j,\nu}$ by the average density $1/2$.
The model with interaction $U$ across all bonds (not only the ones 
within $[1,N]$) shows TLL behavior for $|U| < 2$ with\cite{Haldane} 
\begin{eqnarray}
\label{K}
 K = \left[\frac{2}{\pi}  
 \arccos \left(-\frac{U}{2} \right) \right]^{-1} \; .
\end{eqnarray} 
The Y-junction sketched in Fig.\ \ref{skizze} is described by 
\begin{eqnarray}
\label{hamjunct}
H_{\rm Y} & = & - t_{\rm Y} \sum_{\nu=1}^{3}
 \left( c_{1,\nu}^\dag c_{0,\nu} + \mbox{h.c.} \right) + 
V \sum_{\nu=1}^{3} n_{0,\nu} \nonumber \\* 
&& - t_{\triangle} \sum_{\nu=1}^{3}  \left( e^{i \phi/3} 
c_{0,\nu}^\dag c_{0,\nu+1} +
  \mbox{h.c.} \right)     \; ,
\end{eqnarray}
where we identify the wire indices 4 and 1. 

Our starting point to calculate the linear
response conductance of $\sum_{\nu=1}^{3} H_\nu +
H_{\rm Y} $ is an exact hierarchy of differential flow equations 
for the self-energy $\Sigma^{\Lambda}$ and higher order vertex 
functions (in the presence of the interaction, the leads, and the junction),
with an infrared energy cutoff $\Lambda$ as the flow parameter. 
It is derived using the fRG.
The hierarchy is truncated by neglecting $n$-particle vertices 
with $n>2$, and the 2-particle vertex is parametrized  
by a renormalized nearest neighbor interaction $U^{\Lambda}$. 
This implies that terms of order $U^2$ are only partly taken into 
account and that $\Sigma^{\Lambda}$ is frequency independent. 
The important spatial dependence of $\Sigma^{\Lambda}$, 
is however fully kept. The details of this procedure are given in 
Refs.\ \onlinecite{VM1} and  \onlinecite{Tilman}.
At the end of the flow (at $\Lambda=0$), the self-energy 
can be regarded as an effective, $N$-dependent 
impurity potential $\Sigma$ with non-vanishing matrix elements 
$\Sigma_{j,j}$ and $\Sigma_{j,j+1}$, where $j$ is restricted to 
the interacting region. 
Due to the symmetry of the junction the matrix elements are the same 
for the three TLL wires. 
We here focus on the zero temperature case
for which the flow equations can numerically be solved for 
up to $N=10^7$ lattice sites.\cite{VM1,Tilman}  
Generically $\Sigma_{j,j}$ and $\Sigma_{j,j+1}$ oscillate around an 
average value with an amplitude that decays slowly 
with increasing distance from the junction.

\begin{figure*}[htb]
\begin{center}
\includegraphics[width=0.37\textwidth,clip]{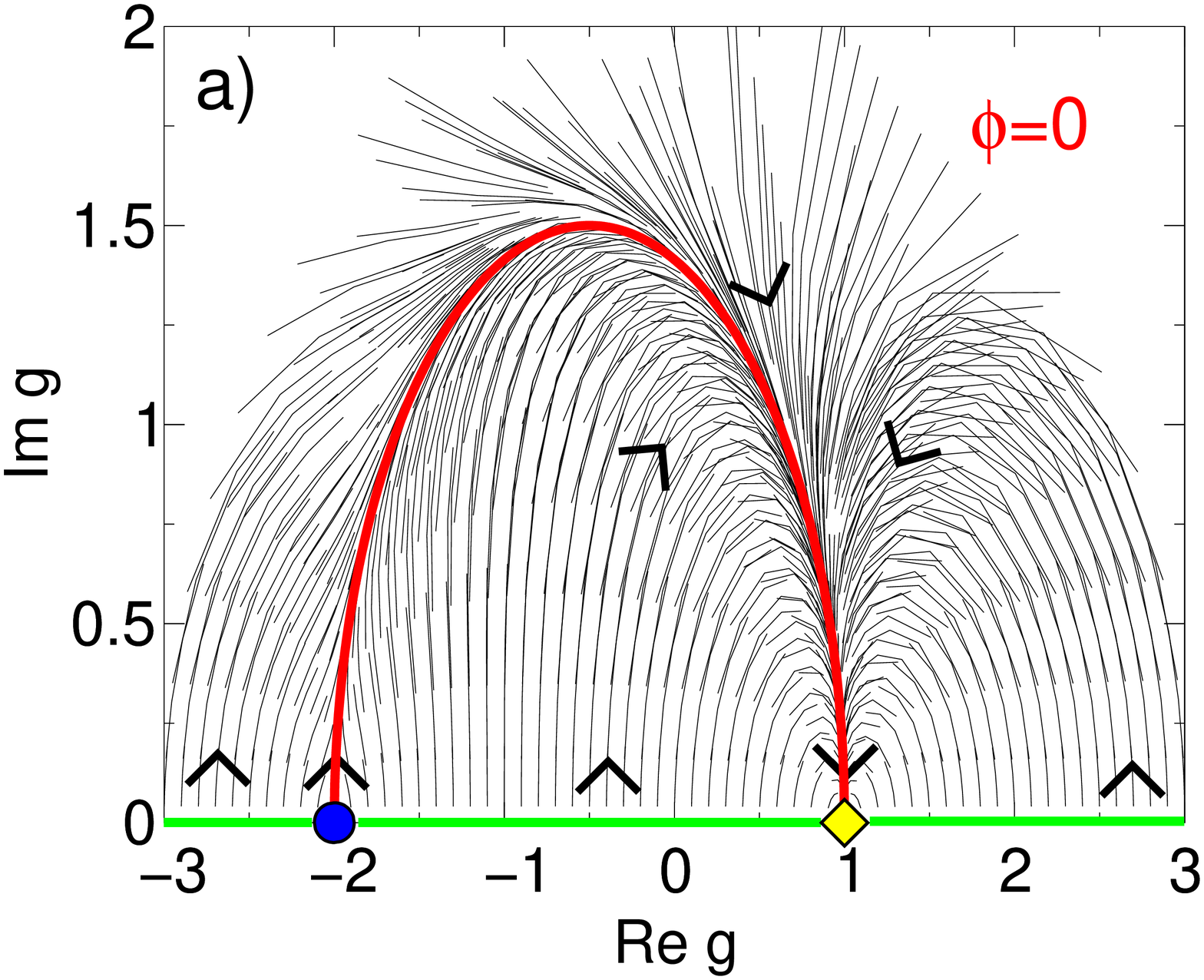}
\includegraphics[width=0.37\textwidth,clip]{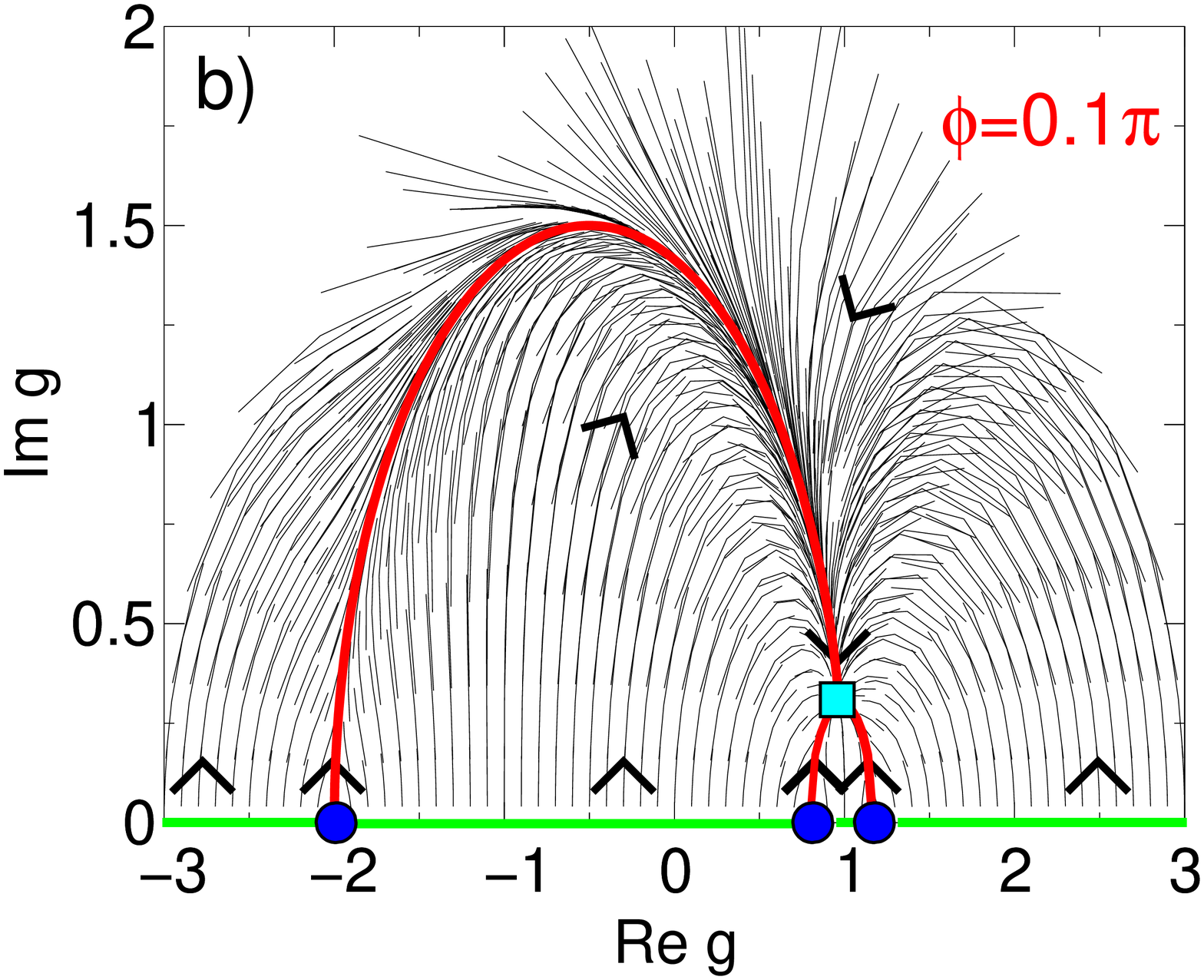}\\
\includegraphics[width=0.37\textwidth,clip]{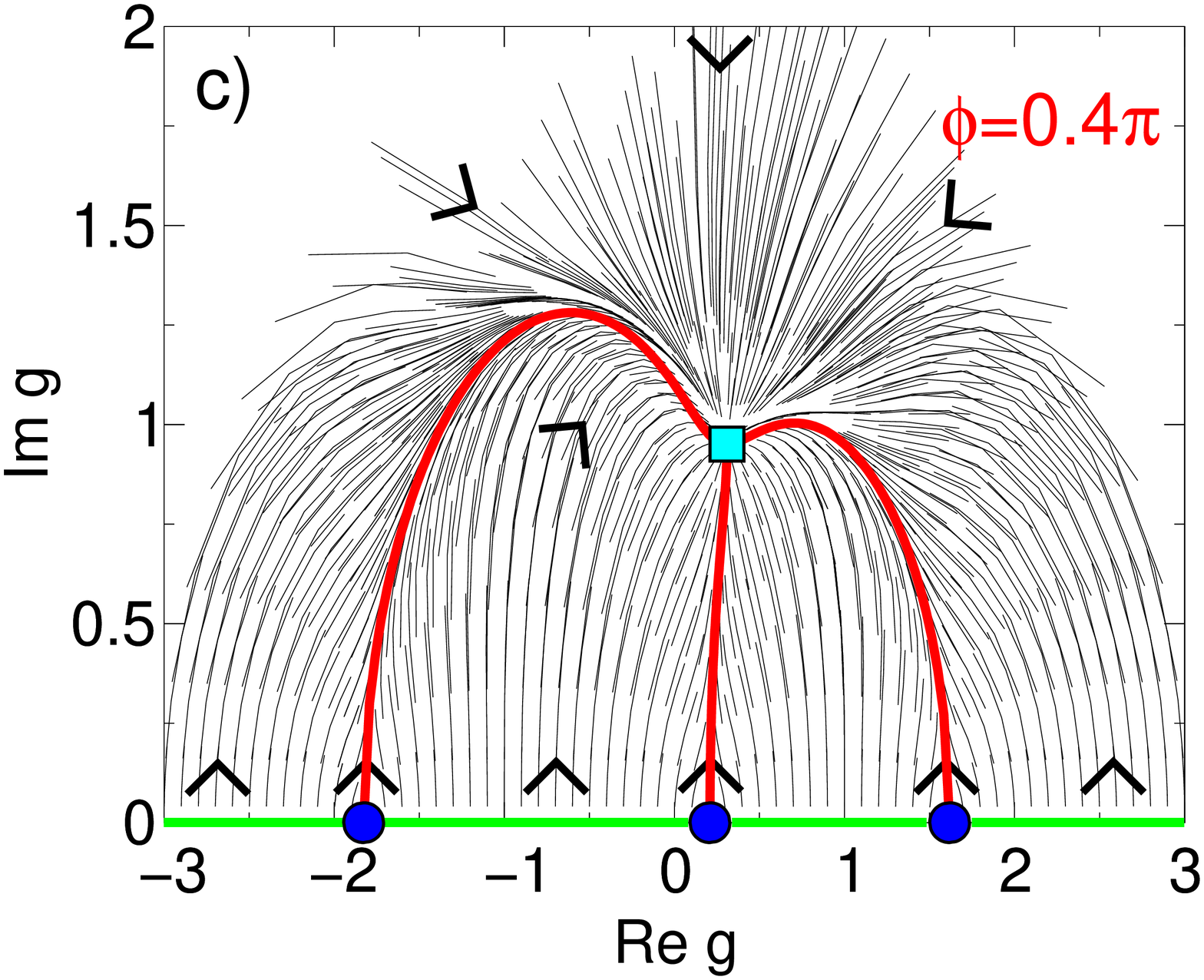}
\includegraphics[width=0.37\textwidth,clip]{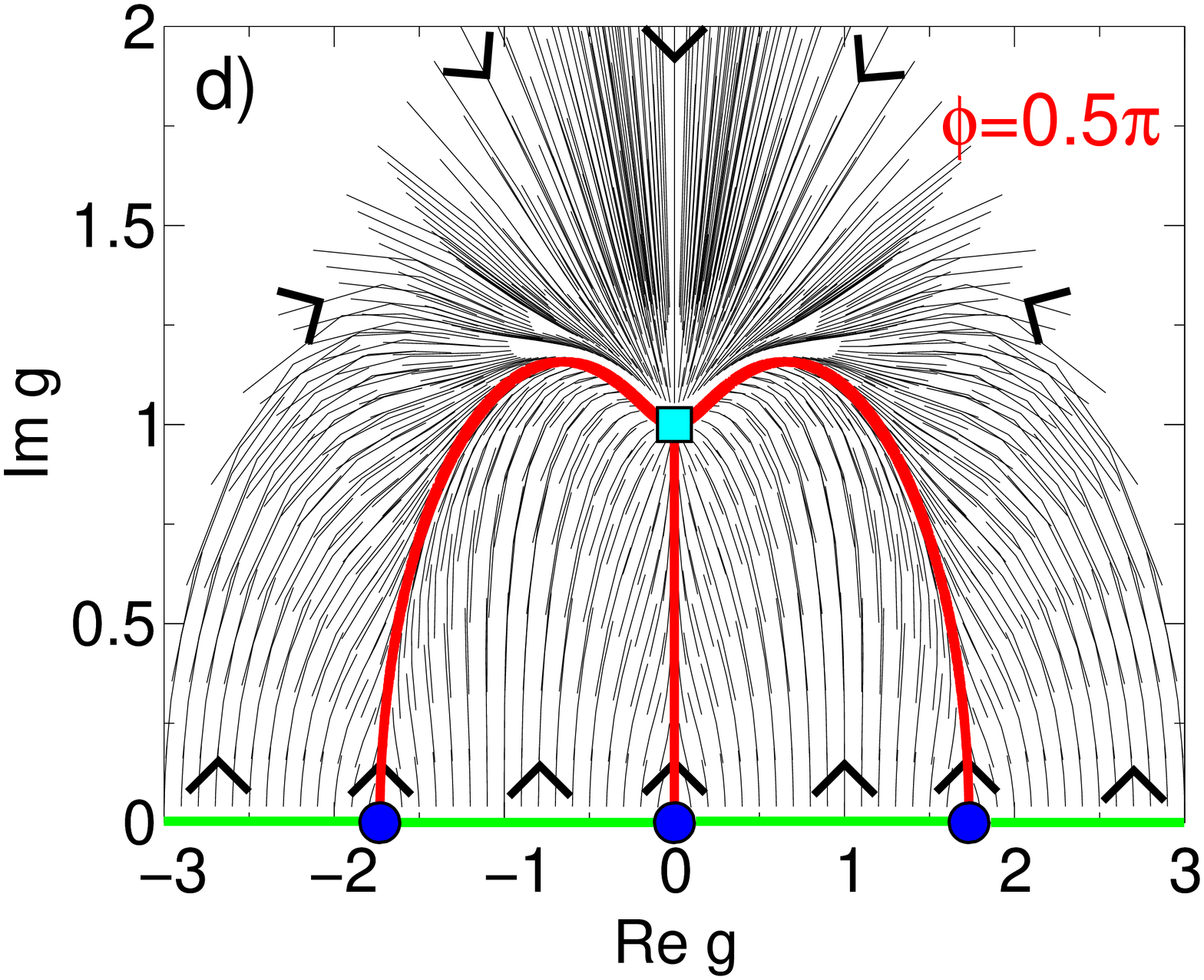}
\end{center}
\vspace{-0.6cm}
\caption[]{(color online) Flow of $g$ for different $\phi$. 
Arrows indicate the direction for $U<0$. For $U>0$ it is reversed. 
For details see the text. 
\label{phiallatrak}}
\end{figure*}

The conductance $G_{\nu,\nu'} = |t_{\nu,\nu'}|^2$ can be calculated from the 
effective transmission $t_{\nu,\nu'}$ (at the chemical 
potential $\mu=0$), which in turn can be expressed in terms of real space 
matrix elements of one-particle Green functions of the system. 
Using single particle scattering theory,\cite{Tilman} 
the conductance can be written as
\begin{eqnarray} 
\label{trans}
G_{\nu,\nu'} = \frac{4 \left( \mbox{Im} \, g \right)^2 
\left| e^{-i \phi} - g \right|^2}{\left|
  g^3-3 g+ 2 \cos{\phi} \right|^2} \; ,
\end{eqnarray}
with $g=(-V-t_Y^2 \, \mathcal G_{1,1})/|t_{\triangle}|$. The Green 
function $\mathcal G$ is obtained by considering $\Sigma$ as an effective
potential for a single wire setting $t_Y=0$ and $\mathcal G_{1,1}$
denotes its diagonal matrix element taken at site $j=1$. 
It is evaluated at energy $\varepsilon +i0$ with $\varepsilon \to 0$.
Eq.~(\ref{trans}) holds
if $\nu,\nu'$ are in cyclic order.  $G_{\nu',\nu}$ follows by replacing 
$\phi \to - \phi$. For symmetry reasons we only have to consider
$0 \leq \phi \leq \pi/2$.  In particular  Eq.~(\ref{trans})
can be applied for $U=0$ and shows that the conductance can be 
parameterized by the flux and a {\it single complex} parameter $g$. 
Via the flow of the self-energy $\mathcal G_{1,1}$ for $U \neq 0$ 
develops an additional dependence on ($t_Y,t_{\triangle},V$), $U$, and
most importantly on $N$. 
The energy scale $\delta_N=\pi v_F/N$ (with the 
Fermi velocity $v_F$) is a natural infrared cutoff of our 
problem.\cite{Tilman} 
To obtain a comprehensive picture of the low-energy physics 
we investigate the flow of $g$ as a function of 
$\delta_N$ and use Eq.\ (\ref{trans}) to calculate $G_{\nu,\nu'}$ for a
given $g$. In Fig.\ \ref{phiallatrak} the flow of $g$
is shown for different $\phi$ (and $U=-1$). Each line is 
obtained for a fixed set of junction  parameters
($t_Y,t_{\triangle},V$)  with $N$ as a variable. 
As $\mbox{Im} \, g$ has the opposite sign of $\mbox{Im} \,
{\mathcal G}_{1,1} <0$ it is restricted to positive values.

Eq.\ (\ref{trans}) allows for {\it three} special conductance situations
which turn out to be the FPs of the flow. On the real axis
(green in Fig.~\ref{phiallatrak}),
the conductance $G$ {\it vanishes} except at special, $\phi$-dependent  
points. They are given by the crossings of the real axis
with the set ${\mathcal C}(\phi)$ 
(red in Fig.~\ref{phiallatrak}), on which the reflection 
$R=1-|t_{\nu,\nu'}|^2-|t_{\nu',\nu}|^2$ takes a local minimum. For 
$\phi=0$,  ${\mathcal C}(0)$ is given by the circle 
$(\mbox{Re}\, g +1/2)^2+(\mbox{Im}\, g)^2 =(3/2)^2$. For 
$\phi > 0$, ${\mathcal  C}(\phi)$ has a more complex analytical 
form not presented here. 
At the crossings the conductance is {\it symmetric} with $G=4/9$. They  
are located at $g=-2\cos{(\phi/3)}$, $2 \cos{([\pi\pm \phi]/3)}$ and  
in Fig.~\ref{phiallatrak} are indicated as blue circles and a yellow 
diamond. As a peculiarity of $\phi=0$ on ${\mathcal C}(0)$ one finds 
$G=4/9$ for all $\mbox{Im}\, g$, not only $\mbox{Im}\, g=0$.  
A situation with $G=0$ is also reached for $|g|\to \infty $. 
For $0 < \phi \leq 1/2$ and $g = e^{i \phi}$ 
Eq.\ (\ref{trans}) yields $G_{\nu,\nu'}=1$ and $G_{\nu',\nu}=0$. At
this point, indicated by cyan squares in Fig.\
\ref{phiallatrak}, {\it maximal asymmetry} of the conductance 
is achieved.     

We next discuss the FP scenarios depicted in Figs.\
\ref{phiallatrak}~a)-d). For $U \neq 0$ the general form of the 
flow diagrams is {\it independent} of the absolute value and sign of $U$. 
In Figs.~\ref{phiallatrak} a)-d) results for  $U=-1$ are shown.
At $\phi=0$ we find two  ``perfect junction'' FPs
with $G=4/9$ and a line of ``decoupled chain'' FPs
with $G=0$ (green line). For all $U<0$ the ``perfect junction'' FP at
$g=1$ (yellow diamond) is the only stable FP.  
All trajectories are attracted towards ${\mathcal C}(0)$ and reach 
this FP following ${\mathcal C}(0)$.  
For all $U>0$ it turns unstable and the line of ``decoupled chain''  FPs 
is stable. In addition the ``perfect junction'' FP at $g=-2$ (blue circle) 
is stable. The {\it basin of attraction} of $g=-2$ is given 
by ${\mathcal C}(0)$. Increasing $\phi$ from 0 the ``perfect
junction'' FP at 
$g=1$ splits up into {\it three} FPs -- the two ``perfect junction'' FPs
at $g=2 \cos{([\pi\pm \phi]/3)}$ 
and the  ``maximal asymmetry'' FP at $g=e^{i \phi}$. 
For all $U<0$ the latter is 
the {\it only} stable FP but becomes unstable for $U>0$. 
A third ``perfect junction'' FP is located at $g=-2\cos{(\phi/3)}$.
For all $U>0$ each of the  three  ``perfect junction'' FPs has a 
{\it basin of attraction} given 
by one of the three parts of ${\mathcal C}(\phi)$ which are 
separated by the  ``maximal asymmetry'' FP. 
In addition for $U>0$ the line of ``decoupled chain''  FPs is stable.
This scenario holds up 
to $\phi=\pi/2$, at which  the  ``perfect junction'' FPs are at 
$g=\pm \sqrt{3}$, $g=0$, and the ``maximal asymmetry'' FP is 
at $g=i$. For $\phi=\pi/2$ and $U>0$ there is a single trajectory 
that runs along the imaginary axis to infinity (leading to $G=0$) 
and does not bend back towards the real axis as all other trajectories 
at $U>0$ do. In the mapping of the complex plane onto the Riemann 
sphere the $g=\infty$ FP (north pole) is part of the projected line of
``decoupled chain'' FPs and shows the same stability properties and
scaling dimension.

To obtain the scaling dimensions of the FPs we generically 
analyze the scaling of $G_{\nu,\nu'}$ as a function of $\delta_N$ with respect 
to the FP conductance which is $1$, $4/9$, or $0$ 
depending on the FP studied. 
For sufficiently large $N$ (in some cases up to $10^5$ sites are
required) we find power-law scaling close to all 
FPs. In cases were 
the FPs are not stable this scaling does not present the asymptotic 
behavior, but exponents smaller than 0 can still be read off at 
intermediate $N$. Approaching the ``perfect junction'' FPs 
along $\mathcal C(\phi)$ for $\phi>0$ Eq.~(\ref{trans}) yields 
$G_{\nu,\nu'}-4/9 \propto \mbox{Im} \, g$, i.e.~$G_{\nu,\nu'}-4/9$ and 
$\mbox{Im} \, g$ scale with the same exponent. The scaling dimensions 
of the ``perfect junction'' FPs at $\phi=0$ cannot be read off  
from the conductance as $G=4/9$ on $\mathcal C(0)$ and we use the 
scaling of $\mbox{Im} \, g$.

\begin{figure}[tbh]
\begin{center}
\includegraphics[width=0.34\textwidth,clip]{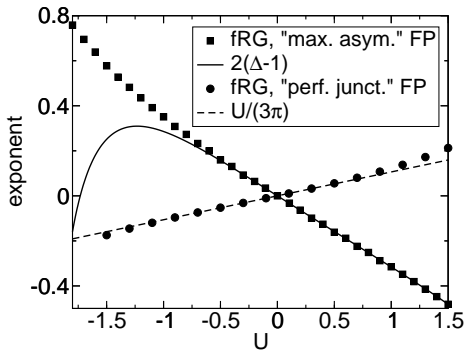}
\end{center}
\vspace{-0.6cm}
\caption[]{Scaling exponents close to FPs.\label{expmafp}}
\end{figure}

The scaling exponent of the ``maximal asymmetry'' FP 
is {\it independent} of the flux $\phi >0$ and the direction from 
which it is approached ($U<0$) or in which it is left ($U>0$). 
Its $U$-dependence is shown 
in Fig.~\ref{expmafp} (squares) and agrees
to leading order in $U$ with the prediction $2(\Delta-1)$ 
(solid line),\cite{Claudio} with $\Delta=4K/(3+K^2)$ and
$K$ given in Eq.~(\ref{K}). In our
scheme terms of order $U^2$ are only partly included 
and we cannot expect agreement to higher order. 
The non-monotonic behavior of  $2(\Delta-1)$ is not reproduced 
and in our approximation the ``maximal asymmetry'' FP stays 
attractive for all $U<0$. For small
to intermediate $|U|$ with TLL parameters $1/2 \leq K
\leq 3/2$ we confirm the conjecture of 
Ref.\ \onlinecite{Claudio}.

For the line of ``decoupled chain'' FPs we find for all $\phi$
the same scaling exponent $\beta_s$, as found for 
the {\it single impurity} problem close to the respective 
``decoupled chain'' FP, applying the fRG.  Its dependence on $U$ is 
shown in Ref.~\onlinecite{Tilman} and agrees  to leading 
order with  $2(1/K-1)$. 

As discussed above the $\phi=0$ ``perfect junction'' FP at 
$g=1$ [yellow diamond in Fig.\ \ref{phiallatrak} a)] 
has properties different from those of the other ``perfect junction'' FPs 
(blue circles in Fig.\ \ref{phiallatrak}). 
The scaling exponent $\gamma$ of the latter FPs is independent of 
$\phi$ and shown in Fig.~\ref{expmafp} (circles). To leading order 
it is given by $\gamma \approx U/(3 \pi)$ (dashed line). 
This form does {\it not} coincide with the expansion of any of 
the above $K$-dependent exponents [after using $K \approx 1 - U/\pi$; 
see Eq.\ (\ref{K})]. We thus find a {\it new} scaling dimension. 
Since higher order terms are only partly included
we cannot determine the functional dependence of $\gamma$ on $K$.
To derive such an expression presents a challenge for methods that do
not require approximations in the strength of the interaction. 
For the $\phi=0$ ``perfect junction'' FP at $g=1$  
the scaling exponent (of $\mbox{Im}\, g$) is,  up to our numerical
accuracy, equal to $- \gamma$. For junction parameters
that initially do not fall onto  ${\mathcal C}(0)$, but are close to
it, one can read off an exponent from the conductance that describes how
${\mathcal C}(0)$, with $G=4/9$, is approached ($U<0$) or left ($U>0$). 
It is equal to the fRG exponent $\beta_w$ of the ``perfect chain'' FP in 
the {\it single impurity} problem. Its $U$-dependence is presented 
in Ref.\ \onlinecite{Tilman} and agrees with  
$2(K-1)$ to leading order. 

The appearance of stable FPs with symmetric conductance $G=4/9$ 
at $U>0$ is a surprising result. Even though time reversal symmetry 
is explicitly 
broken at $\phi >0$, due to correlations the conductance of systems 
with parameters on $\mathcal C(\phi)$ behaves as if time 
reversal symmetry is restored. It is remarkable that 
close to the line of ``decoupled chain'' FPs the relative 
difference $|G_{\nu,\nu'}-G_{\nu',\nu}|/(G_{\nu,\nu'}+G_{\nu',\nu})$ scales as
$ \delta_N^{\beta_s/2}$ and thus vanishes if $U>0$. This implies 
that $G_{\nu,\nu'}$ and $G_{\nu',\nu}$ {\it become equal} faster than they go 
to zero. In that sense also on this line of FPs and thus for {\it all} 
junction parameters time reversal symmetry is restored if $U>0$.  

Using the fRG we determined the complete renormalization group flow
for a TLL Y-junction pierced by a magnetic flux.
Besides uncovering a new type of low-energy physics we 
demonstrated the power of our approximation scheme. Usually  
field theoretical models are used to investigate the transport 
properties of inhomogeneous TLL applying methods that are specific 
to such models. This way {\it exact} results for either fairly simple 
geometries (single impurity\cite{Fendley}) or restricted parameter 
regimes\cite{Claudio} were obtained. Our method can be 
applied to {\it microscopic models} with {\it arbitrary junction 
parameters} and provides results for the conductance 
that are accurate for small to intermediate $|U|$. It can also be used 
to study transport on {\it intermediate} and {\it large} energy
scales.\cite{Tilman} Furthermore, the technique can be applied 
to investigate the transport through more complex networks of TLLs
that might become important in future nano-electronic applications. 
   
We thank C.~Chamon, S.~Andergassen, T.~Enss, W.~Metzner, 
and Th. Pruschke for discussions. V.M. and K.S. are grateful
to the Deutsche Forschungsgemeinschaft (SFB 602) for 
support.


\begin{thebibliography}{*}


\bibitem{KS} For a review see
K.~Sch\"onhammer in {\it Interacting Electrons in Low 
 Dimensions,} Ed.: D.~Baeriswyl, Kluwer Academic Publishers 
 (2004); cond-mat/0305035.      
\bibitem{LutherPeschel} A.~Luther and I.~Peschel, 
Phys.~Rev.~B {\bf 9}, 2911 (1974).
\bibitem{Mattis} D.~Mattis, J.~Math.~Phys.~{\bf 15}, 609 (1974).
\bibitem{KaneFisher} C.~Kane and M.~Fisher, Phys.~Rev.~Lett. 
{\bf 68}, 1220 (1992); Phys.~Rev.~B {\bf 46}, 15233 (1992).
\bibitem{Fendley} P.\ Fendley,  A.~Ludwig, and H.~Saleur, 
Phys.~Rev.~Lett. {\bf 74}, 3005 (1995). 
\bibitem{MatveevGlazman} D.~Yue, L.~Glazman, and K.~Matveev, 
Phys.~Rev.~B {\bf 49}, 1966 (1994).
\bibitem{VM1} V.~Meden {\it et al.,} 
Phys.~Rev.~B {\bf 65},  045318 (2002); J.\ of Low Temp.\ Physics {\bf
  126}, 1147 (2002); S.~Andergassen {\it et al.,} Phys.\ Rev.\ B 
  {\bf 70}, 075102 (2004).   
\bibitem{Tilman} T.~Enss {\it et al.,} to appear in Phys. Rev. B
  (2005).
\bibitem{Fuhrer} M.~Fuhrer {\it et al.,} Science {\bf 288}, 494 (2000).
\bibitem{Terrones} M.~Terrones {\it et al.,} Phys.~Rev.~Lett.~{\bf
    89}, 075505 (2002).
\bibitem{Nayak} C.~Nayak {\it et al.,} Phys.~Rev.~B {\bf 59}, 15694
  (1999); S.~Lal, S.~Rao, and D.~Sen, {\it ibid.} {\bf 66},
  165327 (2002); S.~Chen, B.~Trauzettel, and R.~Egger, 
  Phys.\ Rev.\ Lett.\ {\bf 89}, 226404 (2002). 
\bibitem{Claudio} C.~Chamon, M.~Oshikawa, and I.~Affleck,
  Phys.\ Rev.\ Lett.\ {\bf 91}, 206403 (2003).
\bibitem{Claudioprivate}M.~Oshikawa, C.~Chamon, and I.~Affleck, in
  preparation.
\bibitem{Haldane} F.D.M.~Haldane, Phys.~Rev.~Lett.~{\bf 45}, 1358
(1980). 
\end{thebibliography}
\end{document}